\documentstyle[12pt]{article}
\date{}
\begin{document}

\title{Systematics of the Light and Strange Baryons and
the Symmetries of QCD}

\author{L. Ya. Glozman$^{1,2}$ and D.O. Riska$^1$}
\maketitle

\centerline{\it $^1$Department of Physics, University of Helsinki,
00014 Finland}

\centerline{\it $^2$Alma-Ata Power Engineering Institute,
480013 Alma-Ata, Kazakhstan}

\setcounter{page} {0}
\vspace{1cm}

\centerline{\bf Abstract}
\vspace{0.5cm}

The spectra of the nucleons and the
strange hyperons are well described by a
harmonic confinement potential for the constituent
quarks and an SU(3) flavor-symmetric
interaction mediated by the
pseudoscalar octet
that is associated with the hidden
approximate chiral symmetry of QCD.
The spectrum is formed
of multiplets
of $SU(6)_{FS}\times U(6)_{conf}$, which are mixed
by
the octet interaction.
With 4 matrix elements of the octet mediated interaction
determined by
the lowest mass splittings in the nucleon spectrum
the resulting mass formula predicts the whole spectrum
with an accuracy of $\sim$ 5\%.
\\

PACS numbers 12.39.-x, 12.39.Fe, 12/40.Yx, 14.20.Gk

Submitted to Physical Review Letters

\noindent
\small
Research Institute for Theoretical Physics, University of Helsinki,\\
Preprint HU-TFT-94-48, Dec. 2, 1994\\
LANL hep-ph 9412231

\newpage
\normalsize

	The absence of nearby parity partners to the
lowest states in the spectra of the nucleons and the strange
hyperons shows that
the (approximate) chiral
symmetry of QCD is realized in the
hidden (Nambu-Goldstone) mode at low excitation energy.
This hidden mode of chiral symmetry is
associated with the existence of
the nonet of light pseudoscalar Goldstone
bosons (mesons) and constituent quarks.
The "chiral" pseudoscalar  octet (the $\eta'$ decouples because
of the $U(1)$ anomaly [1]) will mediate interactions between the
constituent quarks. Even
a schematic treatment of this octet mediated interaction
can explain the fine structure of the baryon spectrum in the
light and strange flavor sectors
at the 30\% level under the assumption
that the gross features of the spectra are determined
by a harmonic confining interaction between the
constituent quarks [2].
We here show that when the full orbital structure of the baryon states
is taken into account this model provides a very satisfactory description
of the spectra of the nucleon, $\Delta$-resonance and the
$\Lambda$-hyperon by which all the
resonance energies are predicted to within $\sim$5\%
of the empirical values when
the 4
basic radial integrals of the octet interaction
are extracted
from the lowest mass splittings
in the nucleon spectrum.\\

The simplest representation of the interaction
that is mediated by the octet of pseudoscalar Goldstone
bosons is [2]

$$H_\chi\sim -\sum_{i<j}V(\vec r_{ij})
\vec \lambda^F_i \cdot \vec \lambda^F_j\,
\vec
\sigma_i \cdot \vec \sigma_j.\eqno(1)$$
Here the $\{\vec \lambda^F_i\}$:s are flavor $SU(3)$ matrices and the
$i,j$ sums run over the constituent quarks. The form of this
interaction is an immediate generalization of the spin-spin
component of the pseudoscalar (pion) exchange interaction,
$-\sum_{i<j}g^2\vec\tau_i\cdot\vec\tau_j\vec\sigma_i\cdot\vec\sigma_j
\left[\delta(\vec r_{ij})-\mu^2 exp(-\mu r_{ij})/4\pi r_{ij} \right]
/12m_i m_j$, with a "smeared" $\delta$-function term
($m_i , m_j$ denote the
quark masses). The smearing is related to
the finite size of the constituent quarks and the pseudoscalar
mesons.
The associated tensor component
of the interaction plays an important role
for the the small spin-orbit
splitting of the baryon spectrum, but not for its
main systematics.\\

An important phenomenological reason for considering the
chiral field interaction as the main source of the fine
structure of the baryon spectra in the light flavor sectors
is that it explains the different ordering
of the lowest positive and negative
parity resonances in the
spectra of
the nucleon and the $\Sigma$ hyperon
on the one hand and the $\Lambda$
hyperon on the other.
This feature has proven hard to explain in terms of the
color magnetic interaction [3] that
has been assumed to be the main cause of the fine structure
of the spectrum in
most previous work on the baryon spectra that has employed
the constituent quark model [4,5].
A second such reason is that the pseudoscalar exchange
interaction (1) naturally explains the absence
of the strong spin-orbit interaction that would be
expected to be associated with a gluon exchange model [3].\\

In Tables 1 and 2 we list the spectra of the
nucleons, $\Delta$-resonances and $\Lambda$-hyperons
and their most natural
symmetry classification according to the
group structure $SU(3)_F\times SU(2)_S \times U(6)_{conf}$.
The notation is that of the Elliott scheme as
suggested in ref. [6]. In this notation N is the number of
quanta in the state, which characterizes the $U(6)$ multiplet
of the 3-quark system with harmonic quark-quark interactions.
The Elliott symbol ($\lambda \mu$) characterizes the $SU(3)$
harmonic oscillator symmetry
and $L$ is the orbital angular momentum. The permutational
symmetry of the states is indicated by
$[f]_X$, where f is
the list of boxes in the successive rows of the
corresponding Young patterns.
Similarly the permutational
symmetry of the
flavor-spin $SU(6)_{FS}$, flavor $SU(3)_F$
and spin $SU(2)_S$ states are denoted by corresponding
Young patterns $[f]_{FS},[f]_F$ and $[f]_S$.
The explicit expressions for the wavefunctions are given
in ref. [6].
In the tables we also list the contributions to the
energies caused by the chiral field interaction (1)
as linear combinations of
radial integrals of the interaction potential $V(\vec r)$,
which determine the fine structure splitting according
to the chiral field interaction (1) in first order
perturbation theory. These radial
integrals are defined as

$$P_{nl}=<\varphi_{nlm}(\vec r_{12})
|V(\vec r_{12})|\varphi_{nlm}(\vec r_{12})>, \eqno(2)$$
where the $|\varphi_{nlm}(\vec r_{12})>$ represents
the harmonic oscillator wavefunction
with n excited quanta.\\

If the confining interaction between two constituent
quarks i, j is taken to have the harmonic oscillator form

$$V_{conf}(\vec r_{ij})=V_0+
\frac{1}{6}m\omega^2 (\vec r_i-\vec r_j)^2,\eqno(3)$$
where $m$ is the mass of the constituent quark
and $\omega$ the
angular frequency of the oscillator, the Hamiltonian for the
3-quark state becomes

$$H_0=\sum_{i=1}^3 \frac{\vec p_i^2}{2m}-
\frac{\vec P_{cm}^2}{6m}+\frac{1}{6}m\omega^2
\sum_{i<j}^3(\vec r_i-\vec r_j)^2+ 3V_0.\eqno (4).$$
Here $\vec P_{cm}$ denotes the momentum of the whole baryon.
For simplicity we have
assumed here that the constituent u,d and s quarks
all have equal mass $m$.
The exact eigenvalues and eigenstates to
the Hamiltonian
(4) are then

$$E=(N+3)\hbar\omega+3V_0,\eqno(5a)$$
$$\Psi=|N(\lambda\mu)L[f]_X[f]_{FS}[f]_F[f]_S>.\eqno(5b)$$
The totally antisymmetric color state
$[111]_C$, which is common to all the states, is
suppressed in (5b). By the Pauli principle $[f]_X=[f]_{FS}$.\\

The full Hamiltonian is the sum of the confining Hamiltonian (4)
and the chiral field interaction (1), which causes the
fine structure of the spectrum. When the latter is treated
in first order perturbation theory the total mass of the
baryon states takes the form

$$M=M_0+N\hbar\omega+\delta M\chi.	\eqno(6).$$
Here $M_0=\sum_{i=1}^3 m_i+3(V_0+\hbar\omega)$
and $\delta M_\chi$ is
the fine structure correction $<\Psi|H_\chi|\Psi> $.
In the Tables the fine structure
correction is given explicitly in terms of radial integrals
of the form (2). The interaction (1) is diagonal in states
of definite orbital angular momentum $L$ and
good $[f]_{FS}$, $[f]_F$, $[f]_S$ symmetry and thus
does not cause configuration mixing in first order perturbation
theory. The associated tensor interaction, expected to be weak as
mentioned above, does however mix states with equal $J$ and
flavor symmetry.\\

The baryon states listed in Tables 1, 2 are completely
determined by 4 radial integrals, as long as the
spin orbit splitting
within the multiplets is neglected. Hence one may proceed
phenomenologically by extracting these and the oscillator
parameter $\hbar\omega$ from the lowest mass splittings in
the $N\Delta$ sector and then predict all the other nucleon
and strange hyperon states, as shown below. Beyond that an
explicit model for the interaction potential $V(r)$ in (1)
at short range would be required.\\

The  oscillator parameter $\hbar\omega$ and the 4 integrals
that appear in the two tables are extracted from
the mass differences between the nucleon and the $\Delta(1232)$,
the $\Delta(1600)$ and
the $N(1440)$, as well as the splittings between the nucleon
and the average mass of the
two pairs of states $N(1535)-N(1520)$ and
$N(1720)-N(1680)$.
This procedure yields the parameter values
$\hbar\omega$=157.4 MeV,
$P_{00}$=29.3 MeV, $P_{11}$=45.2 MeV, $P_{20}$=2.7 MeV and
$P_{22}$=--34.7 MeV. Given these values all other excitation energies
of the nucleon, $\Delta$- and $\Lambda$-hyperon spectra are
predicted to within $\sim$ 15\% of the empirical values
where known, and well within the uncertainty limits
of those values. The parameter values above should
be
allowed a considerable uncertainty range in view of
the uncertainty in the empirical values for the
resonance energies. \\

The relative magnitudes and signs
of the numerical parameter values can be readily understood. If
the potential function $V(\vec r)$ is assumed to have the
form of a Yukawa function with a smeared $\delta$-function
term that is positive  at short range $r\le 0.6-0.8$  fm,
as suggested by the pion size $\sqrt{<r_\pi^2>}=0.66$ fm,
one expects $P_{20}$
to be considerably smaller than $P_{00}$ and $P_{11}$,
as the radial wavefunction for the excited S-state has a node,
and as it extends further into region of where the potential
is negative.
The negative value for $P_{22}$ is also natural as the
corresponding wavefunction is suppressed at short range
and extends well beyond the expected
0 in the potential function.
The relatively small value of the oscillator parameter (157.4 MeV)
leads to the empirical value 0.86 fm for the
nucleon radius $\sqrt{<r^2>}=\sqrt{\hbar/m\omega}$
if the quark mass is taken to be 330 MeV,
as suggested by the magnetic moments of
the nucleon.\\

It should be emphasized that
the overall -- sign in the chiral interaction (1)
corresponds to that of the usual pion exchange potential
at short distances, so that the interaction is attractive
in completely symmetric spin-isospin states
and repulsive in
antisymmetrical spin-isospin states.
This argument can be directly extended
to $SU(3)_F$ [2] and hence
symmetrical $FS$ pair states experience an attractive interaction
at short range, whereas antisymmetrical ones experience repulsion.
This
explains why the $[3]_{SF}$ state in the $N(1440)$, $\Delta(1600)$
and
$\Sigma(1660)$ positive parity resonances feels a larger
attractive interaction than the mixed symmetry state $[21]_{SF}$ in the
$N(1535)$,
$\Delta(1700)$
and $\sum(1750)$ resonances. Consequently the masses of the
positive parity states $N(1440)$, $\Delta(1600)$  and
$\Sigma(1660)$ are shifted
down relative to the other ones, which explains the reversal of
the otherwise expected "normal ordering".
The situation is different in the case of the $\Lambda(1405)$ and
$\Lambda(1600)$, as the flavor state of the $\Lambda(1405)$ is
totally antisymmetric. Because of this the
$\Lambda(1405)$ gains an
attractive energy, which is
comparable to that of the $\Lambda(1600)$, and thus the ordering
suggested by the confining oscillator interaction is maintained.\\

The predicted nucleon (and $\Delta$) spectrum,
which in Table 1 is listed up to $N=2$, contains two groups of
nonconfirmed and
unobserved states. These all belong to the
$N=2$-band. The lowest group is the 4 $\Delta$
states around 1675 MeV, one of which plausibly
corresponds to the 1-star $\Delta(1750)$. The predicted
${3\over 2}^+$ and ${5\over 2}^+$ resonances around 1909 MeV
plausibly correspond to the 1- and 2-star resonances
$N(1900)$ and $N(2000)$ respectively.
The predicted $\Lambda$
spectrum contains one unobserved state in the $N=1$ band and 8 in the
$N=2$ band. As these are predicted to lie close to observed states
with large widths their existence is not ruled out.\\

It proves instructive to consider the symmetry structure of the
harmonic confining + chiral octet mediated
interaction (1) model presented here in view of the
highly satisfactory
predictions obtained for the spectra of the nucleon, the $\Delta$
and the $\Lambda$-hyperon.
The symmetry group for the orbital part of a
harmonically bound A quark system is $U(3(A-1))$,
which in the present case reduces to $U(6)$. In the absence
of the fine-structure interaction (1), and with equal
u,d and s-
quark masses, the baryon states would form unsplit
multiplets of the full symmetry group
$SU(6)_{FS}\times U(6)_{conf}$. The chiral interaction (1)
lifts this
degeneracy within the multiplets and is in fact strong enough
to mix members of different multiplets. Thus the N=2 resonance
$N(1440)$ is shifted down below the N=1 resonance $N(1535)$ etc.
When this shifting moves states from adjacent
N-levels close to each other near degenerate parity doublets
appear. The model thus suggests an explicit explanation of
the observed near parity doubling of the spectrum at
high energies, which is however only apparently
accidental [7].\\

It is an empirical fact that the spectra
of the nucleon, the $\Delta$
and the $\Lambda$-hyperon at high excitation
are formed of near parity doublets, the splittings of which
rapidly decrease with energy. In the nucleon spectrum this is
revealed by the shrinkage of the 95 MeV separation in
the $\frac{1}{2}^\pm$ $N(1535)-N(1440)$
parity doublet to only 60 MeV in
the $N(1710)-N(1650)$ doublet. The splitting of the corresponding
$\frac{3}{2}^\pm$ and $\frac{5}{2}^\pm$ parity doublets around
1700 MeV is only 20 MeV and 5 MeV respectively.
In the case of the $\Lambda$-spectrum
the 70 MeV splitting between the $\Lambda(1670)$ and the
$\Lambda(1600)$
$\frac{1}{2}^\pm$ parity doublet
shrinks to only 10 MeV
in the parity doublet $\Lambda(1810)-\Lambda(1800)$. This
gradual transition from a low energy sector with well
separated single states to a near parity doubled spectrum
at high excitation is most naturally explained as a gradual
transition from the hidden realization of the
approximate chiral symmetry of QCD to the explicit
mode [2].
Within the constituent quark model the most natural suggestion
for the appearance of the parity doublets is that the Hamiltonian
that includes confinement (4) and the chiral field interaction (1)
contains a hidden additional symmetry beyond the $SU(6)_{FS}\times
U(6)_{conf}$. The latter is broken by (1). This conjecture is
supported by the relative
insensitivity of predicted spectra to the parameter values used.\\

The approximate chiral symmetry of 3-flavor
QCD in the Wigner-Weyl
mode is the symmetry under independent
rotation of the left and right quark fields in flavor space.
The corresponding
symmetry group is

$$U(3)_L^F\times U(3)_R^F=SU(3)_L^F\times SU(3)_R^F\times
U(1)_A^F\times U(1)_V^F. \eqno(7)$$
 From this product the group $U(1)_A$
should be dropped as the corresponding
symmetry is broken at the quantum
level [8].
The chiral symmetry should in fact be extended to
$SU(3)_L^F\times SU(3)_R^F\times SU(2)^O\times U(1)_V^F$ because
of the spatial rotational invariance ("O")
for the left- and right-handed
quarks.
When the chiral symmetry is realized in the
hidden mode this symmetry group is only
$SU(6)_V^{FO}\times U(1)_V^F = U(6)$.
This is precisely the symmetry of the confining oscillator
Hamiltonian for the 3-quark state. The following conjectures
now suggest themselves: (i) The confining
interaction is
related to the $SU(6)_V^{FO}\times U(1)_V^F$ part of
QCD, and would be harmonic in the chiral limit,
(ii) only 2- and 3-quark systems are confined, since
the required U(3(A-1)) symmetries that
would be required for confinement
of systems
of $A>3$ quarks are not symmetries of the QCD Lagrangian,
(iii) the chiral field interaction (1) causes a
gradual restoration of the full chiral symmetry
$SU(3)_L^F\times SU(3)_R^F\times SU(2)^O\times U(1)_V^F$
and (iv) the emergence of the parity doublets signals the
gradual restoration of explicit chiral symmetry.\\

The model described here has relied on an
interaction potential $V(\vec r)$ in eq.
(1) that is  flavor independent.
A refined version should take into
account the
explicit flavor dependence of the potential function
that is caused by the explicit chiral symmetry breaking in
QCD implied
by the large mass splitting of the pseudoscalar
octet.
That flavor dependence leads e.g. to different values for the
$S$-state matrix elements
$P_{00}^\pi$ and $P_{00}^\eta \simeq P_{00}^K$
of the interactions
that are mediated by the different pseudoscalar mesons.
These may be determined from the octet-decuplet mass
splittings $m_\Delta - m_N = 12 P_{00}^\pi- 2 P_{00}^\eta$
and $m_{\Sigma(1385)}-m_\Sigma = 4P_{00}^\eta+6P_{00}^K$ to be
$P_{00}^\pi$ = 27.6 MeV and
$P_{00}^\eta\simeq P_{00}^K$ = 19.2 MeV. With these values
one may predict the mass splittings within the baryon octet:
$$m_\Lambda-m_N = 6P_{00}^\pi - 6 P_{00}^K+m_s-m_{u,d},\eqno(8a)$$
$$m_\Sigma-m_\Lambda=8P_{00}^\pi-4P_{00}^K-4P_{00}^\eta,\eqno(8b)$$
$$m_\Xi -m_\Sigma=P_{00}^\pi-P_{00}^\eta+m_s-m_{u,d}.\eqno(8c)$$
The last one of these equations gives value
119 MeV for the mass difference between the strange and u,d quarks.
For the $\Lambda-N$ and $\Sigma-\Lambda$ mass differences we
then predict the values 169 MeV and 67 MeV in good agreement
with the corresponding experimental values 176 MeV and 74 MeV
respectively. Moreover the ratio $P_{00}^\pi/P_{00}^K\simeq 1.44 $
agrees well with the ratio of $m_s / m_{u,d}\simeq 1.4$
as expected on the basis of the form of
the $\delta$-function part of the pseudoscalar exchange interaction
(assuming again that $m_u\simeq$ 330 MeV). Note that this explanation
of the octet mass splittings is different from the early
suggestion for explaining it in terms of an interaction of the form
$\vec\sigma_i\cdot\vec\sigma_j V(\vec r)/m_i m_j $, with
$V(\vec r)$ being a flavor independent function [9-12].
The systematics of the spectra
of the $\Sigma$ and $\Xi$ hyperons is predicted to be similar to that
in Table 1, with an obvious assignment of all confirmed
resonances up to 2 GeV. Direct application of
the parameter values used in Tables above will however lead to
an overprediction of 30-130 MeV of the $\Sigma$- and
$\Xi$-resonance energies
because of the neglect there of the flavor dependence of $V(\vec r)$
 and the constituent quark mass differences.
The spectrum of the $\Omega$-hyperon should have the same
structure as that of the $\Delta$-resonance.\\

Quark-quark interactions
that involve the flavor degrees of freedom have been
found to arise in the instanton induced interaction
between the constituent quarks [8]. This interaction
has recently been applied directly to baryon structure [13-15].
It differs
in a crucial aspect from the
pseudoscalar octet mediated interaction (1) in that it
vanishes in flavor symmetric pair states.
As a consequence
it fails to account for the fine structure in the
$\Delta$-spectrum, as exemplified e.g. in the prediction of
the wrong ordering of the $\Delta(1600)$ and the
negative parity pair $\Delta(1620)-\Delta(1700)$ [13].

\vspace{2cm}

{\bf Acknowledgements}
\vspace{0.5cm}

This research has been supported by Academy of Finland Grant \#7635.
Stimulating discussions with F.Coester and V.G.Neudatchin
and valuable correspondence with M.V.Polyakov are acknowledged.

\newpage

{\bf References}
\vspace{0.5cm}

\begin{enumerate}
\item S. Weinberg, Phys. Rev. {\bf D11}. 3583 (1975)
\item L. Ya. Glozman and D.O.Riska, The Baryon Spectrum
and Chiral Dynamics, Preprint HU-TFT-94-47
\item A. DeRujula, H. Georgi and S.L. Glashow, Phys. Rev.
{\bf D12} (1975) 147
\item N. Isgur and G. Karl, Phys. Rev. {\bf D18}, 4187 (1978)
\item N. Isgur and G. Karl, Phys. Rev. {\bf D19}, 2653 (1979)
\item L. Ya. Glozman and E. I. Kuchina, Phys. Rev. {\bf C49}, 1149 (1994)
\item F. Iachello, Phys. Rev. Lett. {\bf 62}, 2240 (1989)
\item G.'t Hooft, Phys. Rev. Lett. {\bf 37}, 8 (1976)
\item Ya. B. Zeldovich and A. D. Sakharov, Yad. Fiz {\bf 4}, 395 (1966)
[Sov. J. Nucl. Phys. {\bf 4}, 283 (1967)]
\item A.D.Sakharov, JETP Lett. {\bf 21}, 554 (1975) [JETP Lett.
{\bf 21}, 258 (1975)]
\item A.D.Sakharov, JETP {\bf 78}, 2112 (1980) [Sov.Phys.JETP
{\bf 51}, 1059 (1980)]
\item A.D.Sakharov, JETP {\bf 79}, 350 (1980) [Sov.Phys.JETP
{\bf 52}, 175 (1980)]
\item A. E. Dorokhov, Yu. A. Zubov and N. I. Kochelev,
Fiz. Elem. Chastits At. Yadra {\bf 23}, 1192 (1992)
[Sov.J.Part.Nucl. {\bf 23}, 522 (1993)]
\item E. V. Shuryak and J. L. Rosner, Phys. Lett. {\bf B218}, 72
(1989)
\item S. Takeuchi and M. Oka, Phys. Rev. Lett. {\bf 66}, 1271 (1991)
\end{enumerate}
\small
\newpage
\centerline{\bf Table 1}

The structure of the nucleon
and $\Delta$ resonance states up to $N=2$,
including 11 predicted unobserved or nonconfirmed
states indicated by question marks.
The predicted energy values (in MeV) are given in the brackets
under the empirical ones.

\begin{center}
\begin{tabular}{|llll|} \hline
$N(\lambda\mu)L[f]_X[f]_{FS}[f]_F[f]_S$
& LS multiplet & average &$\delta M_\chi$\\
&&energy&\\ \hline
$0(00)0[3]_X[3]_{FS}[21]_F[21]_S$ & ${1\over 2}^+, N$ &
939&$-14 P_{00}$\\
&&&\\
$0(00)0[3]_X[3]_{FS}[3]_F[3]_S$ & ${3\over 2}^+, \Delta$ &
1232&$-4 P_{00}$\\
&&(input)&\\
$2(20)0[3]_X[3]_{FS}[21]_F[21]_S$ & ${1\over 2}^+, N(1440)$ &
1440&$-7 P_{00}-7P_{20}$\\
&&(input)&\\
$1(10)1[21]_X[21]_{FS}[21]_F[21]_S$ & ${1\over 2}^-, N(1535);
{3\over 2}^-, N(1520)$ &
1527&$-7 P_{00}+ 5P_{11}$\\
&&(input)&\\
$2(20)0[3]_X[3]_{FS}[3]_F[3]_S$ & ${3\over 2}^+, \Delta(1600)$ &
1600&$-2 P_{00}-2P_{20}$\\
&&(input)&\\
$1(10)1[21]_X[21]_{FS}[3]_F[21]_S$ & ${1\over 2}^-, \Delta(1620);
{3\over 2}^-,\Delta(1700)$ &
1660&$-2P_{00}+6P_{11}$\\
&&(1719)&\\
$1(10)1[21]_X[21]_{FS}[21]_F[3]_S$ & ${1\over 2}^-, N(1650);
{3\over 2}^-,N(1700)$ &
1675&$-2 P_{00}+4P_{11}$\\
&${5\over 2}^-,N(1675)$&(1629)&\\
&&&\\
$2(20)2[3]_X[3]_{FS}[3]_F[3]_S$&${1\over 2}^+,\Delta(1750?);
{3\over 2}^+,\Delta(?)$&1750?&$-2P_{00}-2P_{22}$\\
&${5\over 2}^+,\Delta(?);{7\over 2}^+,\Delta(?)$&(1675)&\\
&&&\\
$2(20)2[3]_X[3]_{FS}[21]_F[21]_S$&${3\over 2}^+,N(1720);
{5\over 2}^+,N(1680)$&1700&$-7P_{00}-7P_{22}$\\
&&(input)&\\
$2(20)0[21]_X[21]_{FS}[21]_F[21]_S$ & ${1\over 2}^+, N(1710)$
&1710&$-{7\over 2}P_{00}-{7\over 2}P_{20}+5P_{11}$\\
&&(1778)&\\
$2(20)0[21]_X[21]_{FS}[21]_F[3]_S$ & ${3\over 2}^+, N(?)
$ &?&$-P_{00}-P_{20}+4P_{11}$\\
&&(1813)&\\
$2(20)2[21]_X[21]_{FS}[21]_F[21]_S$ & ${3\over 2}^+, N(1900?);
{5\over 2}^+,N(2000?);$ &1950?
&$-{7\over 2}P_{00}-{7\over 2}P_{22}+5P_{11}$\\
&&(1909)&\\
$2(20)2[21]_X[21]_{FS}[21]_F[3]_S$ & ${1\over 2}^+, N(?);
{3\over 2}^+,N(?)$&1990?
&$-P_{00}-P_{22}+4P_{11}$\\
&${5\over 2}^+,N(?);{7\over 2}^+,N(1990?)$&(1850)&\\
&&&\\
$2(20)0[21]_X[21]_{FS}[3]_F[21]_S$ & ${1\over 2}^+, \Delta(1910)
$ &1910&$-P_{00}-P_{20}+6P_{11}$\\
&&(1903)&\\
$2(20)2[21]_X[21]_{FS}[3]_F[21]_S$ & ${3\over 2}^+, \Delta(1920);
{5\over 2}^+,\Delta(1905)$ &1912
&$-P_{00}-P_{22}+6P_{11}$\\
&&(1940)&\\ \hline
\end{tabular}
\end{center}

\newpage
\centerline{\bf Table 2}
\vspace{0.5cm}

The structure of the $\Lambda$-hyperon
states up to $N=2$, including
10 predicted unobserved or nonconfirmed states indicated
by question marks. The predicted energies (in MeV)
are given in the brackets under the empirical values.
\begin{center}
\begin{tabular}{|llll|} \hline
$N(\lambda\mu)L[f]_X[f]_{FS}[f]_F[f]_S$
& LS multiplet & average &$\delta M_\chi$\\
&&energy&\\ \hline
$0(00)0[3]_X[3]_{FS}[21]_F[21]_S$ & ${1\over 2}^+, \Lambda$ &
1115&$-14 P_{00}$\\
&&&\\
$1(10)1[21]_X[21]_{FS}[111]_F[21]_S$ & ${1\over 2}^-, \Lambda(1405);
{3\over 2}^-,\Lambda(1520)$ &
1462&$-12 P_{00}+4P_{11}$\\
&&(1512)&\\
$2(20)0[3]_X[3]_{FS}[21]_F[21]_S$ & ${1\over 2}^+, \Lambda(1600)$ &
1600&$-7 P_{00}-7P_{20}$\\
&&(1616)&\\
$1(10)1[21]_X[21]_{FS}[21]_F[21]_S$ & ${1\over 2}^-, \Lambda(1670);
{3\over 2}^-, \Lambda(1690)$ &
1680&$-7 P_{00}+5 P_{11}$\\
&&(1703)&\\
$1(10)1[21]_X[21]_{FS}[21]_F[3]_S$ & ${1\over 2}^-, \Lambda(1800);
{3\over 2}^-,\Lambda(?);$ &
1815&$-2 P_{00}+4P_{11}$\\
&${5\over 2}^-,\Lambda(1830)$&(1805)&\\

&&&\\
$2(20)0[21]_X[21]_{FS}[111]_F[21]_S$ & ${1\over 2}^+, \Lambda(1810)
$&1810&$-6P_{00}-6P_{20}+4P_{11}$\\
&&(1829)&\\
$2(20)2[3]_X[3]_{FS}[21]_F[21]_S$ & ${3\over 2}^+, \Lambda(1890);
{5\over 2}^+,\Lambda(1820)$ &
1855&$-7 P_{00}-7P_{22}$\\
&&(1878)&\\
$2(20)0[21]_X[21]_{FS}[21]_F[21]_S$&${1\over 2}^+,\Lambda(?)$&
?&$-{7\over 2}P_{00}-{7\over 2}P_{20}+5P_{11}$\\
&&(1954)&\\
$2(20)0[21]_X[21]_{FS}[21]_F[3]_S$ & ${3\over 2}^+, \Lambda(?)$&
?&$-P_{00}-P_{20}+4P_{11}$\\
&&(1989)&\\
$2(20)2[21]_X[21]_{FS}[21]_F[3]_S$ & ${1\over 2}^+, \Lambda(?);
{3\over 2}^+,\Lambda (?);$&2020?&
$-P_{00}-P_{22}+4P_{11}$\\
&${5\over 2}^+\Lambda(?);{7\over 2}^+,\Lambda(2020?)$&(2026)&\\
&&&\\
$2(20)2[21]_X[21]_{FS}[111]_F[21]_S$ & ${3\over 2}^+, \Lambda(?);
{5\over 2}^+,\Lambda(?)$&
?&$-6P_{00}-6P_{22}+4P_{11}$\\
&&(2053)&\\
$2(20)2[21]_X[21]_{FS}[21]_F[21]_S$ & ${3\over 2}^+,\Lambda(?);
{5\over 2}^+,\Lambda(2110)$ &2110?
&$-{7\over 2}P_{00}-{7\over 2}P_{22}+5P_{11}$\\
&&(2085)&\\ \hline
\end{tabular}
\end{center}

\end{document}